\font\twelverm=cmr10 scaled 1200    \font\twelvei=cmmi10 scaled 1200
\font\twelvesy=cmsy10 scaled 1200   \font\twelveex=cmex10 scaled 1200
\font\twelvebf=cmbx10 scaled 1200   \font\twelvesl=cmsl10 scaled 1200
\font\twelvett=cmtt10 scaled 1200   \font\twelveit=cmti10 scaled 1200
\font\twelvesc=cmcsc10 scaled 1200  
\skewchar\twelvei='177   \skewchar\twelvesy='60
     
     
\def\twelvepoint{\normalbaselineskip=12.4pt plus 0.1pt minus 0.1pt
  \abovedisplayskip 12.4pt plus 3pt minus 9pt
  \belowdisplayskip 12.4pt plus 3pt minus 9pt
  \abovedisplayshortskip 0pt plus 3pt
  \belowdisplayshortskip 7.2pt plus 3pt minus 4pt
  \smallskipamount=3.6pt plus1.2pt minus1.2pt
  \medskipamount=7.2pt plus2.4pt minus2.4pt
  \bigskipamount=14.4pt plus4.8pt minus4.8pt
  \def\rm{\fam0\twelverm}          \def\it{\fam\itfam\twelveit}%
  \def\sl{\fam\slfam\twelvesl}     \def\bf{\fam\bffam\twelvebf}%
  \def\mit{\fam 1}                 \def\cal{\fam 2}%
  \def\sc{\twelvesc}               \def\tt{\twelvett}
  \def\sf{\twelvesf}
  \textfont0=\twelverm   \scriptfont0=\tenrm   \scriptscriptfont0=\sevenrm
  \textfont1=\twelvei    \scriptfont1=\teni    \scriptscriptfont1=\seveni
  \textfont2=\twelvesy   \scriptfont2=\tensy   \scriptscriptfont2=\sevensy
  \textfont3=\twelveex   \scriptfont3=\twelveex  \scriptscriptfont3=\twelveex
  \textfont\itfam=\twelveit
  \textfont\slfam=\twelvesl
  \textfont\bffam=\twelvebf \scriptfont\bffam=\tenbf
  \scriptscriptfont\bffam=\sevenbf
  \normalbaselines\rm}
     

     
\def\beginlinemode{\endmode
  \begingroup\parskip=0pt \obeylines\def\\{\par}\def\endmode{\par\endgroup}}
\def\beginparmode{\endmode
  \begingroup \def\endmode{\par\endgroup}}
\let\endmode=\par
{\obeylines\gdef\
{}}
\def\singlespace{\baselineskip=\normalbaselineskip}

\def\oneandahalfspace{\baselineskip=\normalbaselineskip
  \multiply\baselineskip by 3 \divide\baselineskip by 2}
\def\doublespace{\baselineskip=\normalbaselineskip \multiply\baselineskip by 2}

\newcount\firstpageno
\firstpageno=2
\footline={\ifnum\pageno<\firstpageno{\hfil}\else{\hfil\twelverm\folio\hfil}\fi}
\def\toppageno{\global\footline={\hfil}\global\headline
  ={\ifnum\pageno<\firstpageno{\hfil}\else{\hfil\twelverm\folio\hfil}\fi}}
\let\rawfootnote=\footnote              
\def\footnote#1#2{{\rm\singlespace\parindent=0pt\parskip=0pt
  \rawfootnote{#1}{#2\hfill\vrule height 0pt depth 6pt width 0pt}}}
\def\raggedcenter{\leftskip=4em plus 12em \rightskip=\leftskip
  \parindent=0pt \parfillskip=0pt \spaceskip=.3333em \xspaceskip=.5em
  \pretolerance=9999 \tolerance=9999
  \hyphenpenalty=9999 \exhyphenpenalty=9999 }
\def\dateline{\rightline{\ifcase\month\or
  January\or February\or March\or April\or May\or June\or
  July\or August\or September\or October\or November\or December\fi
  \space\number\year}}
\def\received{\vskip 3pt plus 0.2fill
 \centerline{\sl (Received\space\ifcase\month\or
  January\or February\or March\or April\or May\or June\or
  July\or August\or September\or October\or November\or December\fi
  \qquad, \number\year)}}
     
     
\hsize=6.5truein
\vsize=8.5truein  
\parskip=\medskipamount
\def\\{\cr}
\twelvepoint            
\doublespace            
\overfullrule=0pt       

\def\title                      
  {\null\vskip 3pt plus 0.2fill
   \beginlinemode \doublespace \raggedcenter \bf}
     
\def\author                     
  {\vskip 3pt plus 0.2fill \beginlinemode
   \singlespace \raggedcenter\sc}
     
\def\affil                      
  {\vskip 3pt plus 0.1fill \beginlinemode
   \oneandahalfspace \raggedcenter \sl}
     
\def\abstract                   
  {\vskip 3pt plus 0.3fill \beginparmode
   \singlespace ABSTRACT: }
     
\def\endtopmatter               
  {\endpage                     
   \body}
     
\def\body                       
  {\beginparmode}               
     
\def\head#1{                    
  \goodbreak\vskip 0.5truein    
  {\immediate\write16{#1}
   \raggedcenter \uppercase{#1}\par}
   \nobreak\vskip 0.25truein\nobreak}
     
\def\subhead#1{                 
  \vskip 0.25truein             
  {\raggedcenter {#1} \par}
   \nobreak\vskip 0.25truein\nobreak}
     
\def\beginitems{
\par\medskip\bgroup\def\i##1 {\item{##1}}\def\ii##1 {\itemitem{##1}}
\leftskip=36pt\parskip=0pt}
\def\enditems{\par\egroup}
     
\def\beneathrel#1\under#2{\mathrel{\mathop{#2}\limits_{#1}}}
     
\def\refto#1{$^{#1}$}           
     
\def\references                 
  {\head{References}            
   \beginparmode
   \frenchspacing \parindent=0pt \leftskip=1truecm
   \parskip=8pt plus 3pt \everypar{\hangindent=\parindent}}

\gdef\refis#1{\item{#1.\ }}                     
     
\gdef\journal#1, #2, #3, 1#4#5#6{               
    {\sl #1~}{\bf #2}, #3 (1#4#5#6)}            

\gdef\refa#1, #2, #3, #4, 1#5#6#7.{\noindent#1, #2 {\bf #3}, #4 (1#5#6#7).\rm} 

\gdef\refb#1, #2, #3, #4, 1#5#6#7.{\noindent#1 (1#5#6#7), #2 {\bf #3}, #4.\rm} 

\def\pr{\journal Phys.Rev., }

\def\prl{\journal Phys.Rev.Lett., }

\def\np{\journal Nucl.Phys., }

\def\annp{\journal Ann.Phys.(N.Y.), }

\def\endreferences{\body}

\def\endpage                    
  {\vfill\eject}
     
\def\endpaper                   
  {\endmode\vfill\supereject}

\def\ref#1{Ref.~#1}                     
\def\Ref#1{Ref.~#1}                     
\def\[#1]{[\cite{#1}]}
\def\cite#1{{#1}}
\def\(#1){(\call{#1})}
\def\call#1{{#1}}
\def\taghead#1{}
\def\frac#1#2{{#1 \over #2}}
\def\half{{\frac 12}}

\def\12{{1\over2}}


\catcode`@=11
\newcount\r@fcount \r@fcount=0
\newcount\r@fcurr
\immediate\newwrite\reffile
\newif\ifr@ffile\r@ffilefalse
\def\w@rnwrite#1{\ifr@ffile\immediate\write\reffile{#1}\fi\message{#1}}

\def\writer@f#1>>{}
\def\referencefile{
  \r@ffiletrue\immediate\openout\reffile=\jobname.ref%
  \def\writer@f##1>>{\ifr@ffile\immediate\write\reffile%
    {\noexpand\refis{##1} = \csname r@fnum##1\endcsname = %
     \expandafter\expandafter\expandafter\strip@t\expandafter%
     \meaning\csname r@ftext\csname r@fnum##1\endcsname\endcsname}\fi}%
  \def\strip@t##1>>{}}

\def\citeall#1{\xdef#1##1{#1{\noexpand\cite{##1}}}}
\def\cite#1{\each@rg\citer@nge{#1}}	

\def\each@rg#1#2{{\let\thecsname=#1\expandafter\first@rg#2,\end,}}
\def\first@rg#1,{\thecsname{#1}\apply@rg}	
\def\apply@rg#1,{\ifx\end#1\let\next=\relax
\else,\thecsname{#1}\let\next=\apply@rg\fi\next}

\def\citer@nge#1{\citedor@nge#1-\end-}	
\def\citer@ngeat#1\end-{#1}
\def\citedor@nge#1-#2-{\ifx\end#2\r@featspace#1 
  \else\citel@@p{#1}{#2}\citer@ngeat\fi}	
\def\citel@@p#1#2{\ifnum#1>#2{\errmessage{Reference range #1-#2\space is bad.}%
    \errhelp{If you cite a series of references by the notation M-N, then M and
    N must be integers, and N must be greater than or equal to M.}}\else%
 {\count0=#1\count1=#2\advance\count1 by1\relax\expandafter\r@fcite\the\count0,
  \loop\advance\count0 by1\relax
    \ifnum\count0<\count1,\expandafter\r@fcite\the\count0,%
  \repeat}\fi}

\def\r@featspace#1#2 {\r@fcite#1#2,}	
\def\r@fcite#1,{\ifuncit@d{#1}
    \newr@f{#1}%
    \expandafter\gdef\csname r@ftext\number\r@fcount\endcsname%
                     {\message{Reference #1 to be supplied.}%
                      \writer@f#1>>#1 to be supplied.\par}%
 \fi%
 \csname r@fnum#1\endcsname}
\def\ifuncit@d#1{\expandafter\ifx\csname r@fnum#1\endcsname\relax}%
\def\newr@f#1{\global\advance\r@fcount by1%
    \expandafter\xdef\csname r@fnum#1\endcsname{\number\r@fcount}}

\let\r@fis=\refis			
\def\refis#1#2#3\par{\ifuncit@d{#1}
   \newr@f{#1}%
   \w@rnwrite{Reference #1=\number\r@fcount\space is not cited up to now.}\fi%
  \expandafter\gdef\csname r@ftext\csname r@fnum#1\endcsname\endcsname%
  {\writer@f#1>>#2#3\par}}

\def\ignoreuncited{
   \def\refis##1##2##3\par{\ifuncit@d{##1}%
    \else\expandafter\gdef\csname r@ftext\csname r@fnum##1\endcsname\endcsname%
     {\writer@f##1>>##2##3\par}\fi}}

\def\r@ferr{\endreferences\errmessage{I was expecting to see
\noexpand\endreferences before now;  I have inserted it here.}}
\let\r@ferences=\references
\def\references{\r@ferences\def\endmode{\r@ferr\par\endgroup}}

\let\endr@ferences=\endreferences
\def\endreferences{\r@fcurr=0
  {\loop\ifnum\r@fcurr<\r@fcount
    \advance\r@fcurr by 1\relax\expandafter\r@fis\expandafter{\number\r@fcurr}%
    \csname r@ftext\number\r@fcurr\endcsname%
  \repeat}\gdef\r@ferr{}\endr@ferences}


\let\r@fend=\endpaper\gdef\endpaper{\ifr@ffile
\immediate\write16{Cross References written on []\jobname.REF.}\fi\r@fend}

\catcode`@=12

\citeall\refto		
\citeall\ref		%
\citeall\Ref		%

\def\a{{\alpha}}
\def\b{{\beta}}

\def\s{\sigma}
\def\half{{1 \over 2}}
\def\ra{{\rangle}}
\def\la{{\langle}}

\def\ih{{i \over \hbar}}

\def\E{{\cal E}}

\def\q{{\bar q}}
\def\f{{\bar f}}
\def\x{{\bar x}}

\def\X{{\bar X}}
\def\Y{{\bar Y}}
\def\D{{\cal D}}
\def\E{{\cal E}}

\def\Tr{{\rm Tr}}

\def\jjh{E-mail address:j.halliwell@ic.ac.uk}



\centerline{\bf Effective Theories of Coupled Classical and Quantum Variables}
\centerline{\bf from Decoherent Histories:} 
\centerline{\bf A New Approach to the Backreaction Problem}

\vskip 0.3in
\author J. J. Halliwell\footnote{$^{\dag}$}{\jjh}
\affil
Theory Group, Blackett Laboratory
Imperial College, London SW7 2BZ
UK
\vskip 0.5in
\centerline {\rm Preprint IC 96--97/45, quant-ph/9705005. May, 1997}
\vskip 0.2in 
\centerline {\rm Submitted to {\sl Physical Review D}}

\abstract    

{We use the decoherent histories approach to quantum theory to
derive the form of an effective theory describing the coupling of
classical and quantum variables. The derivation is carried out for a
system consisting of a large particle coupled to a small particle
with the important additional feature that the large particle is
also coupled to a thermal environment producing the decoherence
necessary for classicality. The effective theory is obtained by
tracing out both the environment and the small particle variables.
It  consists of a formula for the probabilities of a set of
histories of the large particle, and depends on the dynamics and
initial quantum state of the small particle. It has the form of an
almost classical particle coupled to a stochastic variable whose
probabilities are determined by a formula very similar to that given
by quantum measurement theory for continuous measurements of the
small particle's position. The effective theory gives intuitively
sensible answers when the small particle is in a superposition of
localized states (unlike the simple mean field approach of coupling
to the expectation values of the small system).  The derived
effective theory suggests a form of the semiclassical theory even
when the quantum theory of the large system is not known, as is the
case, for example, when a classical gravitational field is coupled
to a quantized matter field, thus offerering a new approach to the
backreaction problem.}

\endtopmatter

\head{\bf 1. Introduction}

What happens when a classical system interacts with a quantum system
in a non-trivial superposition state? Quantum field theory in curved
spacetime is an example of a number of situations where one would
like to know the answer this question. There, the effect of the
quantized matter field on the classical gravitational field is
often assessed using the semiclassical Einstein equations
[\cite{Ros,Mol}]:
$$
G_{\mu \nu} = 8 \pi G \la T_{\mu \nu} \ra 
\eqno(1.1)
$$
The left hand side is the Einstein tensor of the classical
metric field $g_{\mu \nu} $ and the right hand side is the
expectation value of the energy momentum tensor of a quantum
field. 

Although we do not yet have the complete, workable theory of quantum
gravity required to derive an equation like (1.1), on general
grounds it is clear that it is unlikely to be valid unless the
fluctuations in $T_{\mu\nu} $ are small [\cite{For,HaH,KuF}].
Indeed, (1.1) fails to give intuitively sensible results when the
matter field is in a superposition of localized states 
[\cite{PaG,Kib1}]. It is by no means obvious,  however, that we have
to resort to quantum gravity to accommodate non-trivial matter
states. This leads one to ask whether
there exists a semiclassical theory with a much
wider range of validity than (1.1), which gives intuitively
reasonable results for non-trivial superposition
states for the matter field.

The object of the present paper is to derive the form of an
effective theory of coupled classical and quantum variables, in 
some simple models where the quantum theory of the entire system is
known. From there, we can then make a reasonable postulate as to the
form such a theory might take even when the quantum theory of the
variables treated classically is not known.  Of course, a number of
previous authors have attempted either to derive or postulate the
form of theories of coupled classical and quantum variables
[\cite{Ale}]. What is perhaps missing
from most of these earlier approaches is an adequate
characterization of what it means for one of the subsystems to be
effectively classical. Here, we will work in the context of the
decoherent histories approach to quantum theory
[\cite{GeH1,GeH2,Gri,Omn}], where a thorough characterization of
what it means to be classical has been undertaken.
This issue is an involved one, but simply, 
the system must be described by a decoherent
set of histories consisting of the same type of variables at each
moment of time whose probabilities are strongly peaked about
classical equations of motion.

It should be stressed that we do not expect to derive a consistent
theory describing the coupling of {\it fundamentally} classical
variables to quantum variables. Rather, we are looking for the form
of an effective theory in which variables which are the classical
descendents of a (perhaps unknown) quantum theory couple to quantum
variables. In contrast to fundamentally classical variables,
classical descendents of quantum variables always suffer a certain
amount of imprecision, partly due to their quantum fluctuations, but
largely as a result of the coarse graining required for decoherence
and hence, to render them effectively classical. This imprecision
feeds into the quantum variables they couple to,  and, as we shall
see, confers some useful features. 

In this paper, we will concentrate on some simple models in
non-relativistic quantum theory. To motivate the discussion,
consider the following system. Suppose we have a large (``to be
classical'') particle with coordinates $X$ linearly coupled to a
small particle with coordinates $x$. Let the action be
$$
S= \int dt \ \left( \half M \dot X^2 + \half m \dot x^2 - \half m
\omega^2 x^2 - \lambda X x \right) 
\eqno(1.2) 
$$ 
Hence the equations of motion are 
$$ 
\eqalignno{ 
M \ddot X + \lambda x &= 0 
&(1.3)\cr 
m \ddot x + m \omega^2 x + \lambda X &= 0 
&(1.4) \cr }
$$

A naive semiclassical approach (the mean field approach), on which
(1.1) is based, involves considering the equation
$$
M \ddot X + \lambda \la x \ra = 0
\eqno(1.5)
$$
together with the Schr\"odinger equation for the state of the small
system with $X(t)$ as an external classical source. However, as
stated above we do not expect (1.5) to have a very wide range of
validity.  

Physically, when a large, classical particle interacts with a small
quantum system, the large particle in some sense 
``measures'' the position of the
small system at each moment of time, and then evolves according to
the measured value. The probability for the large particle to
measure a particular value of $x$ will be determined by the quantum
state of the small system and there will generally be non-zero
probabilities for a wide range of different values of $x$. There is
no reason why the average value, $\la x \ra $, is the one that will
almost always be measured, unless the coupling is very weak, or the
distribution of $x$ is strongly peaked about $\la x \ra $.
Therefore, what we expect in general is an ensemble of trajectories
for the large particle, with a probability for each trajectory
determined by the quantum state of the small particle. In this
paper, we will derive a scheme of this type, using the decoherent
histories approach, in a class of simple models.

We mention in passing that it is possible to proceed differently
from this point and directly write down  a phenomenological scheme
for the coupling of classical and quantum variables using continous
quantum measurement theory [\cite{BLP,BeS,CaM,Dio3,Dio4}].  Such an
approach was considered in  Ref.[\cite{DiH}]. The idea is that in
(1.3), $X$ is treated as a classical variable, and $x$ is replaced
by  a classical stochastic variable $\x (t) $, the probability for
which is given by a standard construction of quantum measurement
theory:
$$
\eqalignno{
p [ \x (t) ] = & \int \D x \D y \ \ \rho^B_0 (x_0, y_0) \ \exp \left( 
- \int dt { ( x - \x )^2 \over 2 \s^2_1 }
- \int dt { ( y - \x)^2 \over 2 \s^2_1 } \right)
\cr
& \times
\exp \left( \ih \int dt \left( \half m \dot x^2 - \half m \omega^2
x^2 - \lambda x X \right) \right)
\cr
& \times
\exp \left( - \ih \int dt \left( \half m \dot y^2 - \half m \omega^2
y^2 - \lambda y X  \right) \right)
&(1.8) \cr }
$$
where $\rho^B_0 (x_0, y_0)$ is the initial density matrix of the
small quantum system. Therefore, the scheme is to solve the
equations of motion for the large particle
with $\x (t)$ regarded as a classical source,
and then the probability distribution on the trajectories $X(t)$ is
that implied by the probability (1.8). 

The formula (1.8) contains an arbitrary parameter $\s_1 $
representing the imprecision in the continous measurement.
A reasonable estimate as to its value can be made by appealing to
the fact that, as stated above,
$X$ is not fundamentally classical but a classical descendent of a
quantum variable.
It therefore has intrinsic imprecision, which limits the
precision with which it can carry out ``measurements'' of the small
particle. The parameter $\s_1$ ought therefore to be approximately
determined given the size of the fluctuations in $X$ and the nature
of the coupling between $X$ and $x$ [\cite{DiH}].

The scheme we derive in this paper turns out to be very closely
related to the phenomenological scheme presented in Ref.[\cite{DiH}]
(although it is not exactly the same). Furthermore, it yields a
definite value for the parameter $\s_1$.

We will use the decoherent histories approach to quantum theory
[\cite{GeH1,GeH2,Gri,Omn,Hal2}].
In this approach, the primary focus is on the probabilities for
a set of histories of a closed system:
$$
p(\a) = \Tr \left( P_{\a_n} (t_n) \cdots P_{\a_2} (t_2) 
P_{\a_1} (t_1) \rho P_{\a_1}
(t_1) \cdots P_{\a_n} (t_n) \right)
\eqno(1.9)
$$
Here, the $P_{\a} (t) $ are projection operators in the Heisenberg
picture,
$$
P_{\a} (t) = e^{ \ih H t } P_{\a} e^{ - \ih H t }
\eqno(1.10)
$$
They are exhaustive and exclusive, which means, respectively,
$$
\sum_{\a} P_{\a} = 1, \quad P_{\a} P_{\b} = \delta_{\a \b} P_{\a}
\eqno(1.11)
$$
The projection operators describe the possible properties the system
may have at each moment of time. In this paper we are mainly
interested in histories characterized by imprecisely specified
positions, in which case the projectors have the form,
$$
P_{\a} = \int_{\a} dx | x \ra \la x |
\eqno(1.12)
$$
where the integral is over some interval on the real axis labeled by
$\a$. In practice it is often more convenient to work with so-called
Gaussian projectors,
$$
P_{\x} = {1 \over ( 2 \pi \s^2)^{\half} }\int_{-\infty}^{\infty} 
dx \exp \left( - { ( x - \x )^2 \over 2 \s^2} \right)
\eqno(1.13)
$$
which are only approximately exclusive.

Probabilities generally cannot be assigned to sets of histories
unless there is negligible interference between them. The measure of
interference between any pair of histories is the decoherence functional,
$$
D(\a, \a') = \Tr \left( P_{\a_n} (t_n) \cdots P_{\a_2} (t_2) 
P_{\a_1} (t_1) \rho P_{\a_1'}
(t_1) \cdots P_{\a_n'} (t_n) \right)
\eqno(1.14)
$$
When the condition of (approximate) decoherence is satisfied,
$$ 
D (\a, \a')  \approx 0 \ \quad {\rm for}  \quad \a \ne \a'
\eqno(1.15)
$$ 
the interference
between histories is negligible and probabilities obeying the
probability sum rules may be assigned using the formula (1.9).
The decoherence condition is typically only satisfied for histories
that are coarse grained, {\it i.e.}, histories for which the
projections ask only very limited questions. 

A number of recent papers have used the decoherent histories
approach to discuss the emergence of classical behaviour in simple
particle models [\cite{GeH2,Har3,Hal3,HZ1,DoH}].  The decoherent
histories approach is perhaps the most useful approach to this
problem primarily for the following reason. When we say, in the
context of quantum theory, that a particle exhibits almost classical
behaviour, we mean that the probability that it is found at a
sequence of imprecisely specified positions at a sequence of times
exists, and furthermore, that  this probability is peaked about
classical equations of motion [\cite{Har1,GeH2}]. Hence to talk
about classical properties of a point particle, we need to talk
about the histories of imprecisely specified positions. 

A commonly used coarse graining procedure to ensure that 
histories of position are decoherent is to couple to a thermal
environment. We therefore consider projections at each moment of
time of the form
$$
P_{\a} = P^A_{\a} \otimes I^{\E}
\eqno(1.16)
$$
where $P^A_{\a}$ denotes imprecise position projections
for the particle, and
$I^{\E} $ denotes the identity on the environment.
Using this basic set-up, a number of recent papers have shown that
for a thermal environment of sufficiently high temperature,
there exist decoherent histories of imprecisely specified position
[\cite{GeH2,DoH}].
Furthermore, the probabilities for histories are then strongly peaked
about classical equations of motion with dissipation, with thermal
fluctuations about them [\cite{GeH2,Hal3}]. 
If the particle is sufficiently massive,
the effect of the thermal fluctuations is very small, and its
behaviour may therefore be said to be effectively classical.

Given, therefore, this characterization 
of what it means for a particle to be
effectively classical, we may now turn to the main question we are
interested in, which is to determine the form of the effective
equations of motion when the classical particle is coupled to a
small quantum particle. It should be clear that it is very easy to
set up this problem in the decoherent histories approach. We quite
simply couple a small particle in an arbitrary initial state
to the case considered above.
The closed system we consider therefore consists of a large particle
(A) coupled to a small particle (B). The large particle is also
coupled to a thermal environment ($\E$). The projections at each
moment of time are therefore of the form,
$$
P_{\a} = P^{A}_{\a} \otimes I^B \otimes I^{\E}
\eqno(1.17)
$$
where $I^B$ denotes the identity for the small particle. We again
expect decoherence of the histories. The main thing we are
interested in is the probability distribution for the histories of
the large particle: we expect it to be similar to the case above, but 
modified in a way depending on the dynamics of the small particle
and also its initial state (which we leave arbitrary).

Note that, since we are only interested in the effective equations of
motion for the large particle, $A$, we do not consider projections
onto the properties of the small particle, $B$. The histories of the
small particle therefore do not need to be decoherent  (indeed ,the
interesting case is that in which they might exhibit quantum behaviour),
and it is for this reason that we do not couple the small particle
to the environment.

As we shall see, it is easy to set up the expression for the
probabilitity for histories of the large particle. The main issue is
to express the result in a useful and recognizable form.  We shall
show that the effective equations of motion have the form of the
classical equations of motion coupling $X$ to $x$, but with the
small particle variables $x$ replaced by a stochastic c-number $\x
(t) $.  Moreover, the probability distribution for $\x (t) $ is given
by a formula bearing a close resemblance to the probability for a
continuous position measurement in continuous quantum measurement
theory. 

The majority of our results are described in Section 2, where we
consider the simple linear model described above, linearly
coupled to a thermal environment. We compute the probabilities for
histories of the large particle. It has the form of a stochastic
theory in which a classical variable $X$ is coupled to a stochastic
variable $\x (t) $ with a probability distribution for $\x (t)$. 
The distribution of $\x (t) $, in this simple linear model,
essentially reduces to a Wigner function on the initial phase space
data of the small particle (although smeared over a large region of
phase space, so that it is positive). We discuss some properties of
the scheme, and show that the naive semiclassical approximation is
recovered in the limit of very weak coupling.  We also show that if
the small particle is initially in a superposition of localized
states, the large particle ``sees'' one or other of the localized
states, and not the mean position of the entire state.

In Section 3 we demonstrate the connection with quantum theory of
continous measurements. We show that the probability distribution
for $\x (t)$ is closely related to the formula for continuous
quantum measurements, (1.8), and discuss the connection with the
phenomenological scheme of Ref.[\cite{DiH}].

The generalization to non-linear systems with non-trivial couplings
is straightforward, and is  considered in Section 4. Couplings
involving the energy of the small particle in Section 5. We
summarize and conclude in Section 6.

\head {\bf 2. A Simple Linear Model}

We now compute the decoherence functional for a simple linear model.
The model consists of a large free particle linearly coupled to a
small harmonic oscillator (with action Eq.(1.2)), but the large
particle is also coupled to a thermal bath. The large particle could
start out in an arbitrary state, but we are assuming it is almost
classical, so it is most useful to start it out in a state with
almost definite position and momentum. The near--classical behaviour
of the large particle is assured by its coupling to the environment.
The small particle starts out in an arbitrary initial state. We
would like to know how the large classical particle responds to the
presence of the small quantum particle in an arbitrary quantum
state. More precisely, what is the effective description of the
large particle, in terms of the quantum state of the small particle?
In the decoherent histories approach we can quite simply calculate
directly the probability that the large particle will take a
particular trajectory.

\subhead{\bf 2(A). Probabilities for Histories}

After tracing out the thermal bath modes,
the decoherence functional for the model is
$$
\eqalignno{
D [ \X, \Y ] =&
\int \D X \D Y \D x \D y  \ \rho^A_0 (X_0, Y_0) \ \rho^B_0 (x_0, y_0)
\cr
& \times \exp \left( - \int dt { ( X - \X )^2 \over 2 \sigma^2 }
- \int dt { ( Y - \Y )^2 \over 2 \sigma^2 } \right)
\cr
& \times \exp \left( \ih \int dt \left( \half M \dot X^2 - \half M \dot Y^2
\right) - D \int dt ( X-Y)^2 \right)
\cr
& \times \exp \left( 
\ih \int dt \left( \half m \dot x^2 - \half m \omega^2 x^2 - \lambda X
x \right) \right)
\cr
& \times \exp \left(
- \ih \int dt \left( \half m \dot y^2 - \half m \omega^2 y^2 - \lambda Y
y \right)
\right)
&(2.1)\cr }
$$
This formula is an elementary generalization of similar ones used in
Refs.[\cite{GeH2,Hal1,Hal3}]. The integration is over paths 
$X(t), Y(t)$, $x(t), y(t)$ which fold
into the initial density matrices $\rho^A (X_0, Y_0)$, $\rho^B (x_0,
y_0) $ at the initial time, and at the final time, $X =Y$ and $x=y$
are integrated over. We have used Gaussian projections of
width $\s$ to specify the trajectories of the large particle
(although we did not need to do this -- exact projections may have
been used, but this is a bit more awkward [\cite{GeH2}]). 
The influence functional formalism of Feynman and Vernon has been
used to handle the thermal bath [\cite{FeV,CaL}].
The only remnant of
this environment is the term proportional to $ (X-Y)^2$, and the
constant $D$ is given by $D = 2 M \gamma k T / \hbar^2 $. For
simplicity we are working in the limit of high temperature and
negligible dissipation, but these restrictions are easily relaxed.

For macroscopic values of $M$, $T$ and $\gamma$, $ D $ is
exceedingly large, thereby very effectively suppressing
contributions from widely different values of $X$ and $Y$. The
coarse graining scale of $X$ and $Y$ is set by the parameter $\s$,
hence the condition for approximate decoherence is $ D > 1/ \s^2 $
[\cite{DoH,Hal1}]. We are generally interested in histories which
are maximally refined, that is, as fine-grained as possible
consistent with a given standard of approximate decoherence
[\cite{GeH2}]. This means, in this case, that $\s$ is taken to be as
small as possible, which means that it is of order $D^{-\half}$.

The probabilities for histories $\X (t)$, which may now be assigned,
are given by the diagonal elements of (2.1).
Introducing 
$$
Q= \half (X+Y), \quad \xi_1 = X - Y
\eqno(2.2)
$$
the integration over $\xi_1$ may be carried out, and the
probabilities are
$$
\eqalignno{
p[ \X (t) ] = &
\int \D Q \D x \D y \ W_0^A (M \dot Q_0, Q_0 ) \ \rho^B (x_0, y_0 )
\cr
& \times \exp \left( - \int dt { ( Q - \X )^2 \over \sigma^2 }
- {1 \over 4 \hbar^2 \tilde D}
\int dt \left( M \ddot Q + \half \lambda (x+y) 
\right)^2 \right)
\cr
& \times \exp \left( 
\ih \int dt \left( \half m \dot x^2 - \half m \omega^2 x^2 - \lambda Q
x \right) \right)
\cr
& \times \exp \left(
- \ih \int dt \left( \half m \dot y^2 - \half m \omega^2 y^2 -
\lambda Q y \right)
\right)
&(2.3) \cr }
$$
where $\tilde D = D + { 1 / ( 4 \s^2 ) } $.
An integration by parts was performed, in the exponent,
which picks up a
boundary term $ - \ih M \dot Q (0) \xi_1 (0) $ (recall that
$ \xi_1 = 0 $ at the final time). The
integration over $\xi_1 (0)$ then effectively
produces the Wigner transform $W_0^A$ of the initial density matrix 
$\rho^A_0 $ [\cite{GeH2,Hal5}]. 

Eq.(2.3) may be written,
$$
\eqalignno{
p [ \X (t) ] =&
\int \D \q \D Q \  W^A_0 (M \dot Q_0, Q_0) \ w_Q [ \q (t) ] 
\cr
& \times \exp \left( - \int dt { ( Q - \X )^2 \over \sigma^2 }
- {1 \over 4 \hbar^2 \tilde D (1- \eta) }
\int dt \left( M \ddot Q + \lambda \q \right)^2 \right)
&(2.4) \cr }
$$
where
$$
\eqalignno{
w_Q [ \q (t) ] =& \int \D x \D y \ \rho_0^B ( x_0, y_0) 
\ \exp \left( - { \lambda^2 \over 4 \hbar^2 \tilde D \eta } \int dt 
\left( {(x+y) \over 2} - \q \right)^2 \right)
\cr
& \times \exp \left( 
\ih \int dt \left( \half m \dot x^2 - \half m \omega^2 x^2 - \lambda Q
x \right) \right)
\cr
& \times \exp \left(
- \ih \int dt \left( \half m \dot y^2 - \half m \omega^2 y^2 -
\lambda Q y \right)
\right)
&(2.5) \cr }
$$
To achieve the decomposition (2.5) we have effectively
deconvolved the second part of the Gaussian in Eq.(2.4), using the
functional integral generalization of the formula,
$$
\exp \left( - (x-y)^2 \right) = \int dz \ \exp \left(
- { (x-z)^2 \over 1- \eta } - { (y-z)^2 \over \eta } \right)
\eqno(2.6)
$$
This deconvolution is of course not unique, and $\eta$ is an
arbitrary constant parametrizing this non-uniqueness (although
clearly the total probability distribution (2.4) is independent of $\eta$).
This trick turns out to be useful for
smearing the Wigner functions of each particle, thereby rendering them positive
[\cite{Hus,Hal5}].

Written in the form (2.4) the probability distribution  now has a
reasonably natural interpretation. First of all recall that we are
assuming that the Wigner function of the large particle is strongly
peaked about particular values of $Q_0$ and $M \dot Q_0$. Hence in
the absence of the coupling to the small particle, Eq.(2.4)
describes a probability distribution for the large particle strongly
peaked about a single  classical solution with prescribed initial
conditions. The width of peaking about the classical solution is
controlled by the factor $\hbar^2 \tilde D$, which is of order $M
\gamma k T $, and this is typically very small for macroscopic
values of $M$, $\gamma$ and $T$.

With the small particle coupled in, however, there is the
integration over $\q(t)$ together with the weight function (2.5).
In the next section,
we will show that it is closely related to the probabality
distribution for continously measuring the position $q(t)$ of the
small particle. Eq.(2.4) is therefore the sought-after result: it
describes an ensemble of trajectories for the large particle with a
weight depending on the initial conditions and dynamics of the small
particle.

\subhead{\bf 2(B). The Weight Function}

The weight function (2.5) may be further evaluated as follows.
Introduce $ q = \half (x+y)$ and $\xi_2 = x-y $. Then the $\xi_2 $
integral may be done with the result,
$$
\eqalignno{
w_Q [ \q (t) ] =& \int \D q \ W_0^B ( m \dot q_0, q_0 )
\ \exp \left( - { \lambda^2 \over 4 \hbar^2 \tilde D \eta } \int dt 
\left( q - \q \right)^2 \right)
\cr
& \times \delta \left[ m \ddot q + m \omega^2 q + \lambda Q \right]
&(2.7) \cr }
$$
where $W_O^B$ is the initial Wigner function of the small particle.
Now let
$$
q(t) = q_0 \cos \omega t + { \dot q_0 \over \omega} \sin \omega t
+ \lambda \int dt' G(t,t') Q(t') + \delta q (t)
\eqno(2.8)
$$
where $G(t,t')$ is the Green function for the harmonic oscillator, 
and $\delta q (0) = 0 = \delta \dot q (0) $. Then the delta
functional in the functional integration in Eq.(2.7)
becomes $\delta [ m \delta \ddot q (t) ] $, which implies, given the
above initial conditions, that
$\delta q (t) =0 $. All that remain are two
ordinary integrations over $q_0$ and $p_0 = m \dot q_0 $:
$$
\eqalignno{
w_Q [ \q (t) ] = & \int dp_0 dq_0 \ W^B_0 ( p_0, q_0 )
&(2.9) \cr
& \times \exp \left( - {\lambda^2 \over 4 \hbar^2 \tilde D \eta }
\int dt \left( 
q_0 \cos \omega t + { p_0 \over m \omega} \sin \omega t 
+ \lambda \int dt' G(t,t') Q(t')
- \q (t) \right)^2 \right) 
\cr }
$$
This shows that the weight function is in fact a Wigner function
smeared over a region of phase space of size $\Delta$, where
$$
\Delta \ \sim \ { \hbar^2 \tilde D m \omega^2 \over \lambda^2 } 
\ \sim \ \ { M \gamma k T \over \lambda^2 } m \omega^2
\eqno(2.10)
$$
The quantity $M \gamma k T / \lambda^2 $ (divided by time) is a
measure of the fluctuations $ ( \Delta x)^2_{th} $ in the position
of the small particle induced by its coupling to the large particle.
The quantity $ \hbar / ( m \omega ) $ is representative of the
quantum fluctuations $ ( \Delta x)^2_q $ of the small particle.  For
a wide range of choices of the parameters of the model  the induced
thermal fluctuations are much larger than the quantum fluctuations, 
and it follows that $ \Delta > > \hbar $. (This will always be the
case, for example, if the large particle is sufficiently massive).
This means, first of all, that as long  as $\eta$ is not too small,
$w_Q [ \q (t) ]$ is positive, even though the Wigner function is
not, since a smearing of the Wigner function over cells larger in
size than about $ \hbar $ yields a positive distribution function
[\cite{Hus,Hal5}].

More importantly, because the smearing is over a cell size very much
greater than $\hbar$, an effect essentially the same as decoherence
of the {\it small quantum system} is produced. To be precise,
suppose the initial state of the small quantum system consisted of a
superposition of localized wavepackets, {\it e.g.}, coherent states.
Then in the Wigner function, the interference terms between these
wavepackets would appear as terms which rapidly oscillate in phase
space on a scale the size of $\hbar$. It is well known that smearing
the Wigner function over a region much large in size than $\hbar$
strongly suppresses these terms (see Refs.[\cite{PHZ,Zur}], for
example). Therefore, for all practical purposes we may replace the
initial Wigner function with a generally mixed state Wigner function
in which the interference terms between wavepackets has been thrown
away. 

Effectively what is happening here is that the small quantum system
alternatives are approximately decoherent because they are
approximately correlated with the decohered large system alternatives.
A similar phenomenon, in the context of quantum measurement theory,
was noted by Hartle [\cite{Har1,Har2}].

It could be the case, of course, that the interesting  parameters
for the model in a particular application are such that (2.9) is not
in fact positive. Then the weight function can no longer be
interpreted as a probability distribution on $ \q (t) $. However,
what one is ultimately interested in is the probability distribution
for histories of the large particle, Eq.(2.4), is this is positive
by construction, for all choices of parameters.

In this simple linear model, a further simplification may be
obtained by inserting (2.9) in (2.5), and carrying out the
integration over $\q (t) $, with the result,
$$
\eqalignno{
p [ \X (t) ] =&
\int dp_0 dq_0 \ W^B_0 (p_0, q_0 )
\ \int \D Q \  W^A_0 (M \dot Q_0, Q_0) 
\ \exp \left( - \int dt { ( Q - \X )^2 \over \sigma^2 } \right)
\cr & \times
\exp \left( - {1 \over 4 \hbar^2 \tilde D  }
\int dt \left( M \ddot Q + \lambda^2 \int dt' G(t,t') Q(t') 
+ \lambda \left[ q_0 \cos \omega t + { p_0 \over m \omega} \sin \omega t
\right] \right)^2 \right)
\cr
&
&(2.11) \cr }
$$
Loosely speaking, this equation tells us to solve the classical equations
of motion for the small particle in terms of their initial data
$p_0$, $q_0$, and then regard these as stochastic variables with a
probability distribution given by a smeared Wigner function.

Another convenient way of writing the weight function (2.5) is
$$
w_Q [ \q (t) ] = \int \D q \ {\cal W} [ q , Q ] 
\ \exp \left( - { \lambda^2 \over 4 \hbar^2 \tilde D \eta }
\int dt \left( q - \q \right)^2 \right)
\eqno(2.12)
$$
where
$$
\eqalignno{
{\cal W} [ q, Q]
= \int \D \xi_2 & \exp \left( - { i \lambda \over \hbar }
\int dt Q \xi_2 \right)
\cr
& \times 
\exp \left( \ih S_B [q + \half \xi_2 ] - \ih S_B [ q - \half \xi_2 ]
\right) \ \rho_0^B ( x_0, y_0 )
&(2.13) \cr }
$$
where, recall, $ q= \half (x+y) $ and $\xi_2 = x-y $, and $S_B [x]$
is the free action of the small particle. The quantity $ {\cal W}
[q,Q] $ is the Wigner functional, introduced by Gell-Mann and Hartle
[\cite{GeH2}], and is defined, in (2.13), by a functional Wigner transform of
the fine-grained decoherence functional for the small particle
position histories. It is analagous to
the ordinary Wigner function in relation to the density 
matrix [\cite{BaJ}].
Hence, in (2.12), we see that the weight function is a smeared
Wigner functional. Like the Wigner function, the Wigner functional
is not positive in general, but in this case at least, the 
smeared Wigner functional (2.12) is.

\subhead{\bf 2(C). Recovery of the Naive Semiclassical Approximation}

It is also of interest to examine (2.4) or (2.11) in the limit of
very weak coupling, {\it i.e.}, small $\lambda$. One readily finds,
that to lowest order,
$$
\eqalignno{
p [ \X (t) ] =&
\ \int \D Q \  W^A_0 (M \dot Q_0, Q_0) 
\ \exp \left( - \int dt { ( Q - \X )^2 \over \sigma^2 } \right)
\cr & \times
\exp \left( - {1 \over 4 \hbar^2 \tilde D  }
\int dt \left( M \ddot Q + \lambda \la q(t) \ra
\right)^2 \right)
&(2.14) \cr }
$$
where
$$
\la q(t) \ra = \la q_0 \ra \cos \omega t + 
{ \la p_0 \ra \over m \omega} \sin \omega t
+ \lambda \int dt' G(t,t') Q(t') 
\eqno(2.15)
$$

The probability distribution is peaked about the naive semiclassical
equation (1.5), and here we see it emerging in the limit of very
weak coupling. This is not surprising, since weak coupling
corresponds to a very imprecise measurement, and the first thing a
very broad measurement sees is the average value.

Although the naive semiclassical approach is very limited,  the
above considerations show how it might still be useful if used with
discretion for certain types of initial states. Note first of all
that the expression (2.14), the naive semiclassical result, is not
only valid for small $\lambda$. It will also be a valid 
approximation to (2.11) for initial states whose Wigner function is
strongly peaked about the mean values of $p$ and $q$, as is the case
for a coherent state.

Now consider the key case of an
initial state that consists of a superposition of two well-separated
phase space localized states, one localized around $p_1,q_1$, the other
localized around $p_2,q_2$:
$$
| \Psi \ra = \a_1 | \Psi_{p_1 q_1} \ra + \a_2 | \Psi_{p_2 q_2} \ra
\eqno(2.16)
$$
Then the Wigner function for this initial state has the form
$$
W = | \a_1 |^2 W_{p_1 q_1} + | \a_2 |^2 W_{p_2 q_2} + \quad
{\rm interference} \ {\rm terms}
\eqno(2.17)
$$
where $W_{p_1 q_1}$ denotes the Wigner function for the state
$ | \Psi_{p_1 q_1} \ra $, and thus is concentrated around
$p_1, q_1$, and similarly for $p_2, q_2$. Now, the point is that, as
we have argued above, the interference terms in the initial Wigner
function become highly suppressed as a result of the smearing in
Eq.(2.9). Therefore the probability distribution (2.11) has the form
$$
p [ \X (t) ] \approx | \a_1 |^2 p_1 [ \X (t) ] + | \a_2 |^2 p_2 [ \X
(t) ]
\eqno(2.18)
$$
where $p_1 [ \X (t) ] $ denotes the probability distribution (2.4),
but with initial Wigner function $W_{p_1 q_1}$ (for the small system),
and similarly for
$p_2 [ \X (t) ] $. But these Wigner functions are strongly peaked about
their mean values, hence the naive semiclassical expression (2.14)
is valid for $p_1 [\X (t) ] $ and $p_2 [ \X (t) ] $ seperately. The
effective description of the coupled classical and quantum system is
therefore that the classical system follows the equations of motion
$$
M \ddot Q + \lambda \la q (t) \ra_1 = 0
\eqno(2.19)
$$
with probability $ | \a_1 |^2 $, and follows
$$
M \ddot Q + \lambda \la q (t) \ra_2 = 0
\eqno(2.20)
$$
with probability $ | \a_2 |^2$, where 
$$
\la q(t) \ra_1 = \la \Psi_{p_1 q_1} | q(t) | \Psi_{p_1 q_1} \ra
\eqno(2.21)
$$
and similarly for $ \la q(t) \ra_2 $. This is clearly the
intuitively sensible result. Applied undiscerningly, the mean field
equations (1.5) would not give this result. Although here we see
that, given the small amount of insight provided by decoherence, the
mean field equations can be used to good effect.

\head {\bf 3. Continuous Measurement Theory}

The weight function $ w_Q [ \q (t) ] $ turns out to be closely
related to  continuous measurement theory [\cite{BLP,BeS,CaM,Dio3,Dio4}]. 
As we have noted, the interaction of the large particle with the
small one constitutes a continuous but imprecise measurement of the
position of the small particle.
In the quantum theory of continuous measurements,
the probability for measuring a trajectory $\q (t)$ up to an
imprecision $\s_1$ is given by the path integral expression,
$$
\eqalignno{
p [ \q (t) ] = & \int \D x \D y \ \ \rho^B_0 (x_0, y_0) \ \exp \left( 
- \int dt { ( x - \q )^2 \over 2 \s^2_1 }
- \int dt { ( y - \q )^2 \over 2 \s^2_1 } \right)
\cr
& \times
\exp \left( \ih \int dt \left( \half m \dot x^2 - \half m \omega^2
x^2 - \lambda Q x  \right) \right)
\cr
& \times
\exp \left( - \ih \int dt \left( \half m \dot y^2 - \half m \omega^2
y^2 - \lambda Q y  \right) \right)
&(3.1) \cr }
$$
As above, introduce the variables $q = \half (x+y) $, 
$\xi = x-y$, hence,
$$
\eqalignno{
p [ \q (t) ] = & \int \D q \D \xi \ \ \rho^B_0 (q_0 + \half \xi_0, q_0
- \half \xi_0) \ \exp \left( 
- \int dt { ( q - \q )^2 \over \s^2_1 }
- \int dt { \xi^2 \over 4 \s^2_1 } \right)
\cr
& \times
\exp \left( - \ih \int dt \xi 
\left( m \ddot q  + m \omega^2 q +  \lambda Q \right) 
- \ih m \dot q (0) \xi (0)
\right)
& (3.2)\cr }
$$
Comparing with the expression for the weight function (2.5), we see
that it is very similar, if $\s_1$, which is so far arbitrary, 
is taken to be,
$$
\s_1^2 = { 4 \hbar^2 \tilde D \eta \over \lambda^2} \ \sim \ { M
\gamma k T \over \lambda^2 }
\eqno(3.3)
$$
Although note that (3.1), (3.2) differ from (2.5) by the absence of
the term
$$
\exp \left( - \int dt { \xi^2 \over 4 \s_1^2} \right)
\eqno(3.4)
$$
in (2.5). Hence Eq.(2.5) is not {\it exactly} the same as the
continuous measurement formula (3.1). However, we will see that they
are very close.

Carrying out the $\xi$ integration we obtain,
$$
\eqalignno{
p[\q(t)] = &\int \D q \ W_0^B ( m \dot q_0, q_0 )
\cr
\times 
& \exp \left(
- \int dt { ( q - \q )^2 \over \s^2_1 }
- { \s^2_1 \over \hbar^2 } \int dt \left( m \ddot q
+ m \omega^2 q + \lambda Q \right)^2 \right)
&(3.5) \cr }
$$
With the above choice of $\s_1$,
$$
{ \s_1^2 \over \hbar^2} \ \sim \ {M \gamma k T \over \hbar^2
\lambda^2}
\eqno(3.6)
$$
For macroscopic values of $M$, $\gamma$ and $T$, and assuming
$\lambda^2$ is not unusually large, the factor of $\hbar^2$ in the
denominator ensures that $\s_1^2 / \hbar^2 $ is very large. The
second exponential in (3.5) is therefore very close to a delta
function,  Eq.(3.5) is therefore very close in form to the
alternative expression for the weight function (2.7).

Note that the width $\s_1$ of the effective ``measurement'' of
$q$ depends on two things. First of all it depends
on the coupling $\lambda$ and is smaller the larger $\lambda $ is, 
corresponding to the notion that stronger interactions produce more
precise measurements. Secondly, it depends on the combination $M
\gamma k T$, which is a measure of the thermal fluctuations endured
by the large particle as a result of its interaction with the
environment, and hence, is a measure of the precision to within
which the trajectory of the large particle is defined.
$\s_1$ increases with increasing $M \gamma k T $, which is to be
expected, since the precision with which the large particle can
measure the small particle depends on the precision with which the
large particle's properties are themselves defined.

Given the close resemblance to continuous quantum measurement
theory, the question remains, why did we not get {\it exactly} the
formula for continuous quantum measurements? After all, in the
decoherent histories approach it is possible to derive  standard
quantum measurement theory ({\it i.e.}, measurements represented by
exact projection operators at discrete moments of time), under
certain idealized conditions [\cite{Har1,Har2}].  The answer to this
is probably to be found in the nature of the simple model we are
considering and in particular, the couplings between the subsystems.
 For example, in the model considered here, we coupled $x$ to just a
single degree of freedom $X$ (in turn coupled to the environment).
Coupling to a large number of degrees of freedom may lead to the
missing factor, (3.4), in the same way that the coupling of $X$ to a
large environment produces the factor involving $ \int dt (X-Y)^2 $
in Eq.(2.1). Hence it is quite possible that a closer connection
between decoherent histories and continuous quantum measurements
might be found by exploring more general types of models couplings.
This is tangential to the main theme of this paper, so will be
explored elsewhere.

It is perhaps of interest to note that the above result on continuous
measurement can be re-expressed in terms of evolution equations, and
this in fact casts our results in a form very close to the original
mean field equations (1.5) [\cite{DiH}].
For a pure initial state, the probability formula for continuous
meausrements (3.1) has the form $ \la \Psi_{\q} | \Psi_{\q} \ra $,
for a wave function $\Psi_{\q} $, whose path integral representation
is given by ``half'' of (3.1). From this one can define a normalized
state $ | \psi \ra $ whose time evolution is given by the non-linear
stochastic equation,
$$
{d \over dt} | \psi \ra = \left( - \ih H - {1 \over  \s_1^2} 
( \hat q - \la q \ra )^2 \right) | \psi \ra + {1 \over \s_1} 
\left( \hat q - \la q \ra \right) | \psi \ra \eta (t)
\eqno(3.7)
$$
Here $\eta (t)$ is Gaussian white noise whose linear and quadratic
means are,
$$ 
\la \eta (t) \ra_S = 0, \quad \quad
\la \eta (t) \eta (t') \ra_S = \delta (t-t')
\eqno(3.8)
$$
where $\la \ \ra_S$ denotes stochastic averaging.
$ \eta $ is related to the measured variable $\q $ by
$$
\q = \la \psi | \hat q | \psi \ra + \half \s_1 \eta (t) 
\eqno(3.9)
$$
$H$ is the Hamiltonian for the small system (in this case a
harmonic oscillator, with $X(t)$ as an external source).

Hence, to the extent that the semiclassical equations we have
derived are equivalent to continuous quantum measurement,
the new equations that replace the mean field equations 
(1.5) are
$$
M \ddot Q + \lambda \la \psi | q | \psi \ra + 
\half \lambda \s_1 \eta (t) = 0
\eqno(3.10)
$$
where $ | \psi \ra $ evolves according to the stochastic non-linear
equation (3.7). Note that with the value of $\s_1$ is given by
(3.3), the noise term $ \lambda \s_1 \eta (t) $ is independent of
$\lambda$, so remains as $\lambda \rightarrow 0 $, and describes the
thermal fluctuations of the large particle.

The noise term describes fluctuations about the mean. This sort of
modification, in the context of (1.1), has been considered before
[\cite{KuF,HuM}]. More significant is the fact the state
evolves according to (3.7), and it is the properties of this
equation that correspond to the separation of initial superposition
states described in Section 2(C) [\cite{DiH}].

The above scheme was put forward in Ref.[\cite{DiH}] as a
phenomenological model for the coupling of classical and quantum 
variables, and the value of $\s_1$ proposed there on general
physical grounds agrees with the one derived here.

\head {\bf 4. Non-Linear Couplings}

Section 2 concentrated entirely on the case of a free particle
linearly coupled to a harmonic oscillator. Now we show how these
considerations can be extended to more complicated cases. First we
consider the case of a particle in a potential $V(X)$ coupled to a
harmonic oscillator via a coupling of the form, $ g(X) x $. The
decoherence functional (2.1) is therefore replaced by
$$
\eqalignno{
D [ \X, \Y ] =&
\int \D X \D Y \D x \D y  \ \rho^A_0 (X_0, Y_0) \ \rho^B_0 (x_0, y_0)
\cr
& \times \exp \left( - \int dt { ( X - \X )^2 \over 2 \sigma^2 }
- \int dt { ( Y - \Y )^2 \over 2 \sigma^2 } \right)
\cr
& \times \exp \left( \ih \int dt \left( \half M \dot X^2 -V(X) 
- \half M \dot Y^2 + V(Y)
\right) - D \int dt ( X-Y)^2 \right)
\cr
& \times \exp \left( 
\ih \int dt \left( \half m \dot x^2 - \half m \omega^2 x^2 - g(X)
x \right) \right)
\cr
& \times \exp \left(
- \ih \int dt \left( \half m \dot y^2 - \half m \omega^2 y^2 - g(Y)
y \right)
\right)
&(4.1)\cr }
$$
This case is actually handled quite simply using the fact that the
integration over $X$ and $Y$ is strongly concentrated around $X=Y$.
As before, write $ X = Q + \half \xi_1 $ and $ Y = Q - \half \xi_1$.
Then we have 
$$
\eqalignno{
V(X) - V(Y) &= \xi_1 V' (Q) + O (\xi_1^3)
\cr
g(X) x - g(Y) y &= g(Q) (x-y) + \half \xi_1 g'(Q) (x+y) + O (\xi_1^2 )
&(4.2) \cr }
$$
The $\xi_1$ integral is readily done, and we obtain for the
probabilities,
$$
\eqalignno{
p[ \X (t) ] = &
\int \D Q \D x \D y \ W_0^A (M \dot Q_0, Q_0 ) \ \rho^B (x_0, y_0 )
\cr
& \times \exp \left( - \int dt { ( Q - \X )^2 \over \sigma^2 }
- {1 \over 4 \hbar^2 \tilde D}
\int dt \left( M \ddot Q + V'(Q) + \half g' (Q) (x+y) 
\right)^2 \right)
\cr
& \times \exp \left( 
\ih \int dt \left( \half m \dot x^2 - \half m \omega^2 x^2 - g(Q)
x \right) \right)
\cr
& \times \exp \left(
- \ih \int dt \left( \half m \dot y^2 - \half m \omega^2 y^2 - g(Q)
y \right)
\right)
&(4.3) \cr }
$$
This may be written
$$
\eqalignno{
p[ \X (t) ] & = 
\int \D Q \D \q \ W_0^A (M \dot Q_0, Q_0 ) \ w_Q [ \q (t) ]
\cr
& \times \exp \left( - \int dt { ( Q - \X )^2 \over \sigma^2 }
- {1 \over 4 \hbar^2 \tilde D (1- \eta) }
\int dt \left( M \ddot Q + V'(Q) + g' (Q) \q \right)^2 \right)
&(4.4) \cr }
$$
where
$$
\eqalignno{
w_Q [ \q (t) ] =& \int \D x \D y \ \rho_0^B ( x_0, y_0) 
\ \exp \left( - { 1 \over 4 \hbar^2 \tilde D \eta } \int dt ( g'(Q))^2
\left( {(x+y) \over 2} - \q \right)^2 \right)
\cr
& \times \exp \left( 
\ih \int dt \left( \half m \dot x^2 - \half m \omega^2 x^2 - g(Q)
x \right) \right)
\cr
& \times \exp \left(
- \ih \int dt \left( \half m \dot y^2 - \half m \omega^2 y^2 - g(Q)
y \right)
\right)
&(4.5) \cr }
$$
Introducing $ q = \half (x+y)$ and $\xi_2 = x-y $, the $\xi_2 $
integral may be done with the result,
$$
\eqalignno{
w_Q [ \q (t) ] =& \int \D q \ W_0^B ( m \dot q_0, q_0 )
\ \exp \left( - { 1 \over 4 \hbar^2 \tilde D \eta } \int dt ( g'(Q))^2
\left( q - \q \right)^2 \right)
\cr
& \times \delta \left[ m \ddot q + m \omega^2 q + g(Q) \right]
&(4.6) \cr }
$$
This is very similar to the continuous measurement formula (3.1),
(3.2) if we allow the imprecision parameter $\s_1$ to depend on the
external field.

The next more complicated case we consider is that in which the
coupling between the particles is of the form $ g(X) f(x) $.
It is straightforward to show that the probability is then
$$
\eqalignno{
p[ \X (t) ] = &
\int \D Q \D x \D y \ W_0^A (M \dot Q_0, Q_0 ) \ \rho^B (x_0, y_0 )
\cr
& \times \exp \left( - \int dt { ( Q - \X )^2 \over \sigma^2 }
- {1 \over 4 \hbar^2 \tilde D}
\int dt \left( M \ddot Q + V'(Q) + \half g' (Q) (f(x)+f(y)) 
\right)^2 \right)
\cr
& \times \exp \left( 
\ih \int dt \left( \half m \dot x^2 - \half m \omega^2 x^2 - g(Q)
f(x) \right) \right)
\cr
& \times \exp \left(
- \ih \int dt \left( \half m \dot y^2 - \half m \omega^2 y^2 - g(Q)
f(y) \right)
\right)
&(4.7) \cr }
$$
Again this may be cast in the form 
$$
\eqalignno{
p[ \X (t) ] & = 
\int \D Q \D \f \ W_0^A (M \dot Q_0, Q_0 ) \ w_Q [ \f (t) ]
\cr
& \times \exp \left( - \int dt { ( Q - \X )^2 \over \sigma^2 }
- {1 \over 4 \hbar^2 \tilde D (1- \eta) }
\int dt \left( M \ddot Q + V'(Q) + g' (Q) \f \right)^2 \right)
&(4.8) \cr }
$$
where
$$
\eqalignno{
w_Q [ \f (t) ] =& \int \D x \D y \ \rho_0^B ( x_0, y_0) 
\ \exp \left( - { 1 \over 4 \hbar^2 \tilde D \eta } \int dt ( g'(Q))^2
\left( {(f(x)+f(y)) \over 2} - \f \right)^2 \right)
\cr
& \times \exp \left( 
\ih \int dt \left( \half m \dot x^2 - \half m \omega^2 x^2 - g(Q)
f(x) \right) \right)
\cr
& \times \exp \left(
- \ih \int dt \left( \half m \dot y^2 - \half m \omega^2 y^2 - g(Q)
f(y) \right)
\right)
&(4.9) \cr}
$$
Again it is similar to the continuous measurement formula,
but now the variable being measure is not position $x$, but $f(x)$,
as one would expect since it is this that couples to the large
particle. It is not possible to evaluate $w_Q$ any further in this case.

Overall, therefore, although the evaluation of the path integrals is
less explicit in these more complicated case, the general pattern is
the same as in the linear case described in Section 2: the large
particle follows near-deterministic equations of motion but with a
stochastic forcing term due the small particle, whose probability
distribution bears a close resemblance to the formula of
continuous quantum measurement theory.

\head{\bf 5. Coupling to Energy}

Another case of particular interest, especially in connection with
the semiclassical Einstein equations (1.1), is the case in which the
large particle couples to the energy of the small particle. The
considerations of the previous sections apply to this case very
easily. In fact, this case turns out to be somewhat simpler.

Let the Hamiltonian of the total closed system, including large
system (A), small system (B) and environment ({\E}), be
$$
H = H_A + H_{AB} + H_\E + H_{A \E}
\eqno(5.1)
$$
where 
$$
H_{AB} = \lambda g(X) h
\eqno(5.2)
$$
and $h$ is a harmonic oscillator Hamiltonian. $g (X) $ is an
arbitrary function of $X$.
$H_A$, $H_{\E}$ and
$ H_{A \E} $ are as before. Let the initial state of the small
system be written in terms of energy eigenfunctions,
$$
\rho^B = \sum_{E E'} \rho_{E E'} | E \ra \la E' |
\eqno(5.3)
$$
where $ h | E \ra = E | E \ra $. Then, because $h$ commutes with
everything, whenever $H$ operates on $ | E \ra $, $h$ is replaced
by the eigenfunction $E$.

It is then straightforward to see that the probabilities for
histories of the large particle are
$$ 
\eqalignno{
p [ \X (t) ] =& \sum_{E} \ \rho_{EE}
\ \int \D X \D Y \ \rho^A_0 (X_0, Y_0) 
\cr
& \times \exp \left( - \int dt { ( X - \X )^2 \over 2 \sigma^2 }
- \int dt { ( Y - \X )^2 \over 2 \sigma^2 } \right)
\cr
& \times \exp \left( \ih \int dt \left( \half M \dot X^2 - \half M \dot Y^2
\right) - D \int dt ( X-Y)^2 \right)
\cr
& \times \exp \left( 
- \ih \int dt \left( g(X)-g(Y) \right) E \right)
&(5.4) \cr}
$$
The summation over $E$ and $E'$ is diagonal because
$h$ commutes with everything else so the states $|E \ra $ are
preserved under evolution by the total Hamiltonian, and the trace
in the decoherence functional then contains the term 
$ \la E | E' \ra $.

Introducing $Q$ and $\xi$ as before, this becomes,
$$
\eqalignno{
p[ \X (t) ] = & \sum_{E} \ \rho_{EE}
\ \int \D Q \ W_0^A (M \dot Q_0, Q_0 ) 
\cr
& \times \exp \left( - \int dt { ( Q - \X )^2 \over \sigma^2 }
- {1 \over 4 \hbar^2 \tilde D}
\int dt \left( M \ddot Q + g' (Q) E  \right)^2 \right)
&(5.5) \cr }
$$
It is straightforward to then rewrite this in terms of continuous
imprecise measurement of energy. For simplicity take $ g(Q) =
\lambda Q $, then
$$
\eqalignno{
p[ \X (t) ] = & \int d {\bar E} \int \D Q \ W_0^A (M \dot Q_0, Q_0 ) 
\ w ( {\bar E} )
\cr
& \times \exp \left( - \int dt { ( Q - \X )^2 \over \sigma^2 }
- {1 \over 4 \hbar^2 \tilde D (1- \eta)}
\int dt \left( M \ddot Q + \lambda {\bar E}  \right)^2 \right)
&(5.6) \cr }
$$
where
$$
w ( {\bar E} ) = \sum_{E} \ \rho_{EE}
\ \exp \left( - { \lambda^2 \tau \over 4 \hbar^2 \tilde D \eta } (E
- {\bar E} ) \right)
\eqno(5.7)
$$
where $\tau$ is the time duration of the histories. Eq.(5.7) is the
formula for the continous imprecise measurement of energy, using
Gaussian projectors, to a width
of order $ \hbar^2 \tilde D / (\lambda^2 \tau )$. Here the
connection with continous measurements is precise, although this is
clearly due to the simplicity of this particular case.

Of perhaps greater interest is the situation in which the energy
of the small particle couples in a non-trivial way to the large
particle, for example, through a small particle Hamiltonian of the
form
$$
H = f(X) p^2 + g (X) x^2
\eqno(5.8)
$$
This is a more realistic model of the way in which matter couples to
gravity (for example, in cosmology,  a single mode of a scalar field
coupled to the scale factor). This is much more complicated to deal
with and will be treated elsewhere.

\head{\bf 6. Discussion}

We have derived the form of the effective equations of motion for
some simple systems consisting of a large particle coupled to a
small particle, and coupled also to a thermal environment in order
to produce the decoherence necessary for classicality of the large
particle. The resultant effective theory has the form of a classical
variable coupled to a stochastic variable $\x (t) $, where the
probability distribution for the stochastic variable is given by a
certain weight function (most generally, Eq. (4.9)). This weight
function is closely related (although not exactly the same) as the
probability for continous imprecise measurements of the position of
the small particle. In the case of coupling to energy, it is exactly
the same as the continous measurement theory result. 

The weight function has the property that it suppresses the
interference between localized wave packets for the small particle. 
Hence one of the more unsatisfactory features of the naive
semiclassical approximation is avoided, and the intuitively sensible
result that localized wave packet initial states may be treated
separately is restored.

The derived semiclassical theory suggests the form of a possible
semiclassical theory even when the quantum theory of the variables
that are taken to be classical is not known. It is the following:
in the equations of motion for the classical system, which involves
a coupling to the quantum system, replace the quantum variables with
stochastic variables whose probabilities are given by a weight
function of the form (2.5) (or its generalizations). 
The classical variable $Q$ is regarded as an external classicl
source in (2.5) and the path integral is well-defined, even if the
quantum theory of $Q$ is not known.

The only aspect of (2.5) that was inherited from the quantum theory
of the classical variables is the width of the Gaussian, $ \hbar^2
\tilde D^2 / \lambda^2 $. However, we can see from (2.4) that
physically, the factor $\hbar^2 \tilde D $ is a measure of the
precision to within which the trajectories of classical variables
are defined, and we can imagine that this number could be determined
(or at least bounded) by experiment.

A very similar semiclassical scheme (using the continous measurement
formula) was described in Ref.[\cite{DiH}], and the results of this
paper give partial substantiations of that scheme.

We have used the decoherent histories approach to derive effective
field equations, for reasons stated in Section 1. It is, however,
quite possible that other approaches to emergent classicality may be
used, such as the density matrix approach [\cite{JoZ,Zur}], the
quantum state diffusion picture [\cite{GP1,Per,HZ1}], or the hybrid
representation of composite quantum systems [\cite{Dio2,Dio6,Dio7}].
A system similar
to that considered in this paper has been analyzed in the quantum
state diffusion picture by Zoupas [\cite{Zou}], and a simple spin
system by Yu and Zoupas [\cite{YuZ}].

\head{\bf Acknowledgements}

I am very grateful to Lajos Di\'osi and Jason Twamley for useful
conversations.

\references

\def\pr{{\sl Phys. Rev.\ }}
\def\prl{{\sl Phys. Rev. Lett.\ }}
\def\prep{{\sl Phys. Rep.\ }}

\def\np{{\sl Nucl. Phys.\ }}

\def\annp{{\sl Ann. Phys. (N.Y.)\ }}

\refis{Ale} I.V.Aleksandrov, Z.Naturf. {\bf 36A}, 902 (1981);
A.Anderson, \prl {\bf 74}, 621 (1995);
\prl {\bf 76}, 4090 (1996);
W.Boucher and J.Traschen, \pr {\bf D37}, 3522 (1988);
K.R.W.Jones, \prl {\bf 76}, 4087 (1996);
L.Di\'osi, \prl {\bf 76}, 4088 (1996);
I.R.Senitzky, \prl {\bf 76}, 4089 (1996).

\refis{BaJ} N.Balazs and B.K.Jennings, \prep {\bf 104}, 347 (1984),
M.Hillery, R.F.O'Connell, M.O.Scully and E.P.Wigner, \prep {\bf
106}, 121 (1984); 
V.I.Tatarskii, {\sl Sov.Phys.Usp} {\bf 26}, 311 (1983).

\refis{BLP} A.Barchielli, L.Lanz and G.M.Prosperi,
{\sl Il Nuovo Cimento} {\bf 72B}, 79 (1982).

\refis{BeS} V.P.Belavkin and P.Staszewski, {\sl Phys.Rev.} 
{\bf A45}, 1347 (1992).

\refis{CaL} A.O.Caldeira and A.J.Leggett, {\sl Physica} {\bf
121A}, 587 (1983).

\refis{CaM} C.M.Caves and G.J.Milburn, {\sl Phys.Rev.} {\bf A36}, 5543 (1987).


\refis{Dio2} L.Di\'osi, ``A True Equation to Couple Classical and
Quantum Variables'', preprint quant-ph/9510028 (1995),

\refis{Dio3} L.Di\'osi, {\sl Phys.Rev.} {\bf A42}, 5086 (1990).

\refis{Dio4} L.Di\'osi, {\sl Phys.Lett.} {\bf 129A}, 419 (1988).


\refis{Dio6} L.Di\'osi, {\sl Quantum Semiclass.Opt.} {\bf 8}, 309
(1996).

\refis{Dio7} L.Di\'osi, preprint quant-ph/9610037 (1996). To appear
in {\it Fundamental Problems in Quantum Physics}, edited by
M.Ferrero and A. van der Merwe (Kluwer, Denver, 1997).

\refis{DiH} L.Di\'osi and J.J.Halliwell,, ``Coupling Classical and
Quantum Variables using Continous Quantum Measurement Theory'',
Imperial College preprint 96-97/46, quant-ph/9705XX (1997).

\refis{DoH} H.F.Dowker and J.J.Halliwell, {\sl Phys. Rev.} 
{\bf D46}, 1580 (1992).


\refis{FeV} R.P.Feynman and F.L.Vernon, \annp {\bf 24}, 118 (1963).

\refis{For} L.H.Ford, \annp {\bf 144}, 238 (1982).

\refis{GeH1} M.Gell-Mann and J.B.Hartle, 
in {\it Complexity, Entropy 
and the Physics of Information, SFI Studies in the Sciences of Complexity},
Vol. VIII, W. Zurek (ed.) (Addison Wesley, Reading, 1990); and in
{\it Proceedings of the Third International Symposium on the Foundations of
Quantum Mechanics in the Light of New Technology}, S. Kobayashi, H. Ezawa,
Y. Murayama and S. Nomura (eds.) (Physical Society of Japan, Tokyo, 1990).

\refis{GeH2} M.Gell-Mann and J.B.Hartle,
{\sl Phys.Rev.} {\bf D47}, 3345 (1993).


\refis{GP1} N. Gisin and I.C.Percival, {\sl J.Phys.} {\bf A26},
2233 (1993); {\bf A26}, 2245 (1993).

\refis{Gri} R.Griffiths, {\sl J.Stat.Phys.} {\bf 36}, 219 (1984).

\refis{Hal1} J.J.Halliwell, ``Aspects of the Decoherent Histories
Approach to Quantum Theory'',
in {\it Stochastic Evolution of Quantum States in Open Systems and 
Measurement Processes}, edited by
L.Di\'osi, L. and B.Luk\'acs (World Scientific, Singapore, 1994).

\refis{Hal2} J.J.Halliwell,
``A Review of the Decoherent Histories Approach to Quantum Mechanics'',
in {\it Fundamental Problems in Quantum Theory}, 
edited by D.Greenberger and A.Zeilinger,
Annals of the New York Academy of Sciences, Vol 775, 726 (1994).

\refis{Hal3} J.J.Halliwell, {\sl Phys.Rev.} {\bf D48}, 4785 (1993).

\refis{Hal5} J.J.Halliwell, {\sl Phys.Rev.} {\bf D46},  1610 (1992).

\refis{HZ1} J.J.Halliwell and A.Zoupas,
{\sl Phys.Rev.} {\bf D52}, 7294 (1995);
``Post-decoherence density matrix propagator for quantum 
Brownian motion'',
IC preprint 95-96/67, quant-ph/9608046 (1996),
accepted for publication in {\sl Phys.Rev.D} (1997).

\refis{Har1} J.B.Hartle, in {\it Quantum Cosmology and Baby
Universes}, S. Coleman, J. Hartle, T. Piran and S. Weinberg (eds.)
(World Scientific, Singapore, 1991).  

\refis{Har2} J.B.Hartle, in {\it Proceedings of the 1992 Les Houches Summer
School, Gravitation et Quantifications}, edited by B.Julia and J.Zinn-Justin 
(Elsevier Science B.V., 1995)

\refis{Har3} J.B.Hartle, in, {\it Proceedings of
the Cornelius Lanczos International Centenary Confererence},
edited by J.D.Brown, M.T.Chu, D.C.Ellison and R.J.Plemmons
(SIAM, Philadelphia, 1994)

\refis{HaH} J.B.Hartle and G.T.Horowitz, \pr {\bf D24}, 257 (1981).

\refis{HuM} B.L.Hu and A.Matacz, \pr {\bf D51}, 1577 (1995).

\refis{Hus} K.Husimi, {\sl Proc.Phys.Math.Soc. Japan} {\bf 22},
264 (1940).

\refis{JoZ} E.Joos and H.D.Zeh, {\sl Zeit.Phys.} {\bf B59}, 223
(1985).

\refis{Kib1} T.W.B.Kibble, in {\it Quantum Gravity 2: A Second Oxford
Symposium}, edited by C.J.Isham, R.Penrose and D.W.Sciama (Oxford
University Press, New York, 1981).


\refis{KuF} C-I.Kuo and L.H.Ford, \pr {\bf D47}, 4510 (1993).


\refis{Mol} C.Moller, in {\it Les Theories Relativistes de la
Gravitation}, edited by A.Lichnerowicz and M.A.Tonnelat (CNRS,
Paris, 1962).

\refis{Omn} R.Omn\`es, {\it The Interpretation of Quantum Mechanics}
(Princeton University Press, Princeton, 1994);
{\sl Rev.Mod.Phys.} {\bf 64}, 339 (1992), and references therein.

\refis{PaG} D.N.Page and C.D.Geilker, \prl {\bf 47}, 979 (1981).

\refis{PHZ} J.P.Paz, S.Habib and W.Zurek, \pr {\bf D47}, 488 (1993).

\refis{Per} I.C.Percival,  {\sl J.Phys.} {\bf A27}, 1003 (1994).

\refis{Ros} L.Rosenfeld, \np {\bf 40}, 353 (1963).


\refis{YuZ} T.Yu and A.Zoupas, in preparation.

\refis{Zou} A.Zoupas, ``Coupling of Quantum to Classical in the
Presence of a Decohering Environment'', Imperial College preprint (1997).

\refis{Zur} W. Zurek, {\sl Prog.Theor.Phys.} {\bf 89}, 281 (1993);
{\sl Physics Today} {\bf 40}, 36 (1991);
in, {\it Physical Origins of Time Asymmetry}, edited by 
J.J.Halliwell, J.Perez-Mercader and W.Zurek (Cambridge
University Press, Cambridge, 1994).

\endreferences

\end